\documentclass[a4paper,12pt]{article}
\usepackage[cp1251]{inputenc}
\usepackage[english,russian]{babel}
\usepackage{indentfirst}
\usepackage{amssymb,amsmath,amsfonts,amsthm}
\usepackage{graphicx}
\usepackage{setspace}
\usepackage{afterpage}
\usepackage{longtable}
\usepackage{bm}
\usepackage[mathscr]{eucal}
\usepackage{color}
\usepackage{subfigure,wrapfig,floatflt,lscape}%,baxodraw}
\usepackage{bm,cite,mathrsfs,upgreek,cancel,mathabx}
 \nonstopmode
\begin{document}
%\newpage
\begin{center}
{\bf The Magnon Hall Effect as Amplifier and Switch  in Current
Conversion Processes in a Magnetic Insulator/Metal Heterostructure}
\end{center}
\begin{center}
{\bf I.I.Lyapilin}
\end{center}
 M.N.Mikheev Institute of Metal Physics of the UB
RAS, S.Kovalevskaya St., 18, Yekaterinburg, 620108, Russia

lyapilin@imp.uran.ru

 {\bf Abstract}

The transformation process (conversion)of a spin-wave current into a
diffusive spin current mediated by conduction electrons has been
examined within the context of the magnon Hall effect taking place
in a magnetic insulator/non-magnetic metal heterostructure. It has
been demonstrated that the thermal magnon Hall effect can be
controlled through external parameters, such as an applied magnetic
field or a temperature gradient, thereby affecting the current
conversion mechanism. This conversion process can be utilized both
to identify the interaction mechanism behind the magnon Hall effect
and to measure its magnitude. Additionally, the thermal magnon Hall
effect has been found to function as a separator, amplifier, and
switch in current conversion processes.

\section{Introduction}

One of the key challenges in spintronics is still to reduce energy
consumption during data processing, storage, and transmission
without sacrificing performance. A promising alternative for
conventional devices that use electrons to carry information is
magnetic insulators, where information can be encoded with magnetic
excitations.

A magnon is a quantum of collective magnetic excitations in a
magnetically ordered material; it possesses a spin angular momentum
oriented opposite to the local magnetic moment. Consequently, a flow
of magnons, generated by  non-equilibrium magnetic excitations,
constitutes a flow of angular momentum, or a spin current. Magnons
hold the potential for using them as spin information carriers. This
stems from several factors, including their high degree of coherence
and their ability to travel over great distances due to low
spin-wave attenuation. Being bosonic quasiparticles with a spin
equal to $\hbar$, magnons exhibit transport properties that are
governed by Bose statistics, carrying only spin-wave current,
$\mathbf J=\mathbf J_M $. Manipulating such excitations does not
require the physical displacement of electrical charges, allowing
for a significant reduction in Joule heating.

Magnonic spin transport is typically driven by a temperature
gradient because magnons are electrically neutral and do not
interact directly with electric fields. The forces operating on  a
system of neutral quasiparticles arise from statistical factors and
make themselves felt in the macroscopic level.

Effects observed in spintronics are often enhanced at the separation
line (interface) between two media. Consequently, one infallible way
of producing spin currents is to employ heterostructures composed of
two or more distinct compounds. The term "conversion" encompasses
phenomena that turn one kind of current into another (e.g., spin
current $\leftrightarrow$ into charge current). Transport phenomena
involving both charged particles and neutral quasiparticles are
significantly influenced by the geometric phase (Berry phase) that a
system of particles acquires during its evolution. In this context,
the particles' group velocity $\hbar\dot{\bm r}$ gains an anomalous
component $\bm\upsilon_a$ a proportional to the curvature of the
gauge field $\bm \Omega(\bm k) $ \cite{Ber, Xiao}.

Calculations based on the application of the Berry phase have proven
successful in explaining transport phenomena in ferromagnetic
metals, semiconductors, and transition-metal oxides.

The magnon Hall effect is caused by a geometric phase that
dramatically influences the nature of magnon motion. Predicted in
ref.\cite{Kac}, the thermal magnon Hall effect (MHE) was
subsequently observed in a ferromagnetic insulator $Lu_2V_2O_7$
\cite{Onose} and studied in crystals with various lattice geometries
\cite{Owerre2, Mook, Chen}. In the physics of magnetic insulators,
the MHE serves as a valuable tool for investigating excitations
within magnetic insulating textures. The anisotropic
Dzyaloshinskii-Moriya exchange interaction is one of the mechanisms
that gives rise to an anomalous velocity and, consequently, to the
MHE \cite{Dzya}.

The MHE, akin to a system of charged particles, manifests as an
influence on the nature of magnon motion, changing the trajectories
of the magnons through their velocity $\bm\upsilon_a$. Below, we
will take a close look at phenomena where the behavior of magnon
motion under the MHE conditions is particularly evident, focusing
specifically on the current conversion process in a ferromagnetic
insulator/non-magnetic metal heterostructure (FD/NM).

Near the boundaries of the FD/NM structure, the action of MHE causes
a local shift in the nonequilibrium concentration of magnons,
"squeezing" the magnons against one of the structure's boundaries.
Such a situation is equivalent to a change in the local chemical
potential $\mu_m(y)$ of the magnons; hence, the MHE can be
controlled by varying the directions ($\mathbf H$) and ($\nabla T$),
which in turn affects the current conversion process. Thus, the MHE
in the device at hand appears as a spatial separator and amplifier
of current conversion processes.

\section{\bf The Microscopic Hamiltonian} Consider a ferromagnetic
insulator/non-magnetic metal heterostructure (FI/NM). The interface
between the media is assumed to lie in the (xy)-plane, oriented
along the x-axis. Along the z-axis, an external magnetic field
$H=(0,0,H_0 )$ is applied. Heisenberg exchange interactions,
Dzyaloshinskii-Moriya interactions, and easy-axis anisotropy (with
the easy axis aligned with the magnetic field direction) are all
included in the ferromagnetic insulator's Hamiltonian that is given
by:
\begin{equation}\label{g1}
\hat{\mathcal{H}}=-J\sum _{\langle i,j\rangle
}\hat{\mathbf{S}}_{i}\cdot \hat{\mathbf{S}}_{j}+\sum _{\langle
i,j\rangle }\mathbf{D}_{ij}\cdot (\hat{\mathbf{S}}_{i}\times
\hat{\mathbf{S}}_{j})-K\sum _{i}(\hat{S}_{i}^{z})^{2}-g\mu _{B}H\sum
_{i}\hat{S}_{i}^{z}.
\end{equation}
Here, $J$ is the Heisenberg exchange integral, J>0; K is the
easy-axis uniaxial anisotropy constant, $K>0$; $\mathbf{D}_{ij}=
-\mathbf D_{ji}$  is the Dzyaloshinskii-Moriya vector, and <ij>
stands for the summation taken over nearest neighbors.

In certain types of free-inversion center lattices, such as the
kagome or pyrochlore lattices, the Dzyaloshinskii-Moriya
antisymmetric exchange interaction (DMI), which arises from
spin-orbit coupling, acts as an effective vector potential for
neutral quasiparticles known as magnons. Consequently, magnon wave
packets (zones n) acquire an anomalous velocity component,
$\bm\upsilon_{na}= \dot{\bm k}\times\bm\Omega_n(\bm k)$, that is
perpendicular to their thermal velocity, $\bm\upsilon_T=(\partial
\epsilon _{n}(\vec{p})/\partial \vec{p}$. The influence of
$\bm\upsilon_{na}$ alters the motion of the magnon flow generated by
a temperature gradient, $\nabla _{x}T$, causing the magnons to be
deflected toward the y-axis direction.

Due to Berry curvature, a transverse (Hall) magnon current arises
along the y-axis. Integrating the Berry curvature over the Brillouin
zone, accounting for the Bose-Einstein distribution, yields the
magnitude of the magnon Hall  conductivity,$\alpha _{xy}$. Suppose
that the thermal magnon Hall effect (MHE) produces the magnon
current that is oriented along the y-axis, away from the interface.
The transverse magnon current $J_{m}^{y}$ contains two components,
such as a thermoelectric component (arising from the MHE) and a
diffusive component (driven by a gradient in the chemical
potential):
\begin{equation}\label{g2}
J_{m}^{y}=-\alpha _{xy}\frac{\partial T}{\partial x}+\sigma
_m\frac{\partial \mu _{m}}{\partial y},
\end{equation}
where, $\sigma _m$ is the magnonic conductivity; $\mu _{m}$ is the
chemical potential of the magnons; and $-\alpha _{xy}\nabla T$ is
the magnon flow induced by a temperature gradient, which underpins
the longitudinal spin Seebeck effect (SSE). The fact that the
thermal magnon Hall effect (MHE) deflects magnons away from the
interface between media implies that the Hall current is channeled
depthward into the ferromagnet, ($J_m(y)> 0$). To establish a
profile of the magnon chemical potential ($\mu_m (y)$), we employ
the steady-state spin diffusion equation for magnons: $\partial
^2\mu _m/(\partial y)^2=\mu _m/\lambda _m^2$. The latter's solution
decaying into the bulk takes the following form:
\begin{equation}\label{g4}
\mu _{m}(y)=\mu _{m}(0)e^{-y/\lambda_{m}},
\end{equation}
where, $\lambda _m=\sqrt{D_m\tau_m}$ is the spin diffusion length of
magnons; $D_m$ is the diffusion coefficient; and $\tau_m$ is the
magnon relaxation time.

The balance of currents can be used to compute the change in the
chemical potential at the interface (at y=0). Assuming that the
total magnon current at the interface is converted into a spin
current $J_s (y)$  injected into the metal via spin pumping, we
have:
\begin{equation}\label{g5}
J_{m}^{y}(0)=G_{mix}\mu _{m}(0),
\end{equation}
where $G_(mix )$ is the effective spin conductivity of the interface
\cite{Wei}. Plugging the expression for the total current, we derive
an expression that defines the shift in the chemical potential at
the interface:
\begin{equation}\label{g6}
\mu_m(0)=\frac{\lambda_m\alpha _{xy}}{\sigma _m
-\lambda_m\,G_{mix}}\nabla_x T
\end{equation}
The expression for the chemical potential at the interface is
directly proportional to the Berry curvature integral
$\alpha_{xy}\propto\Omega_z(\vec k)$. This analysis indicates that
magnons are driven by the MHE away from the interface, $(\alpha_{xy}
\nabla_x T < 0$. Consequently, a magnon deficiency arises in the
vicinity of the interface: the resulting voltage in the non-magnetic
metal decreases as the magnon chemical potential at the interface
drops below its equilibrium value, ($\mu_m (0)<0$). The depth of
this depletion is exactly correlated with the Dzyaloshinskii-Moriya
interaction magnitude (through $\alpha _{xy}$) and $\nabla T$. Thus,
a non-equilibrium state of magnons emerges in the structure. This
phenomenon is exhibited as a result of a geometric imbalance in the
magnon flows, generated by the thermal magnon Hall Effect (MHE). The
magnon intake (or outflow) rate toward (from away) the interface,
propelled by the Berry curvature, competes with the rate of their
relaxation (or generation) at the metal interface.
\subsection*{\bf Conclusions}
The contribution of the Thermal Magnon Hall Effect (MHE) to the
modification of the longitudinal Seebeck effect is proportional to
the spin diffusion length $\lambda _{m}$. The purer the crystal
(i.e., the higher the value of $\lambda _{m}$), the wider the
depletion region, and the more strongly the TMHE suppresses the
longitudinal Seebeck effect. In this geometry, the magnon Hall
effect functions as an effective thermodynamic pump that alters the
boundary conditions for the spin current.

Since the magnitude of the spin current $\mathbf J_s$ injected from
a ferromagnet (FM) into a non-magnetic material (NM) via spin mixing
is directly proportional to the excess chemical potential of magnons
at the interface, we have $J_s \propto \mu_m(0)$. Consequently, the
diversion of magnons away from the interface must lead to a
significant suppression of the measured spin Seebeck effect signal.
Under a strong Dzyaloshinskii-Moriya interaction (DMI) and great
temperature gradients, the magnon Hall effect can almost "block" the
thermal flow of magnons toward the interface, thereby minimizing
spin accumulation in the metal. Thus, the MHE operates as a
destructive factor for the longitudinal spin Seebeck effect in a
geometry where the deflection of magnon trajectories is directed
away from the interface. If the ferromagnetic layer thickness $L_y$
is comparable to $\lambda _m$, an excess of magnons arises at the
opposite (free) boundary, ($\mu_m > 0$). For a bulk
Dzyaloshinskii-Moriya interaction, this suggests that the sign of
the change in the longitudinal Seebeck effect will be strictly
opposite if the metallic contact is moved to the other side of the
sample.

In structures that include a topological magnonic insulator,  the
impact of the thermal magnon Hall effect (MHE) on current conversion
processes is likely more pronounced. Since the energy gap
($\epsilon_n-\epsilon_m$ ) between magnonic bands in the expression
for Berry curvature $\Omega_{n}(\mathbf{k})$ becomes negligible at
certain spatial points, the Berry curvature must rise in a singular
manner. In this case, the integrated Hall conductivity
$\alpha_{xy}^{m}$ increases significantly compared to that of
conventional ferromagnets, resulting in a substantial excess of
magnons ($\Delta n_0$ ) at the topological insulator/non-magnetic
metal interface \cite{Wang}. A similarly non-trivial scenario is
expected to occur in structures consisting of an antiferromagnetic
insulator and a metal. The coexistence of two or more spin
sublattices in an antiferromagnetic system introduces an internal
degree of freedom, giving rise to magnons with different chiralities
that propagate at distinct velocities. Undoubtedly, this situation
is bound to affect current conversion processes.

This work was carried out within the framework of a state assignment
from the Ministry of Science and Higher Education of the Russian
Federation to the Institute of Metal Physics, Ural Branch of the
Russian Academy of Sciences (project "Spin," state registration
number 122021000036-3).

The author declares no conflict of interest.

\end{document}